\documentclass[11pt]{article}
\usepackage{cite}
\usepackage{amsmath,amsfonts}
\usepackage{graphicx,color}
\usepackage{epsfig,epsf}
\usepackage{hyperref}
\usepackage{bm}

\usepackage{relsize}
\def\Babar{{\mbox{\slshape B\kern-0.1em{\smaller A}\kern-0.1em B\kern-0.1em{\smaller A\kern-0.2em R}}}}

\def\L {\Lambda}

\def\bar {\overline}

\def\bea {\begin{eqnarray}}
\def\eea {\end{eqnarray}}
\def\n {\nonumber}

\def\beq{\begin{equation}}
\def\eeq{\end{equation}}
\def\barr{\begin{array}}
\def\earr{\end{array}}
\def\dis{\displaystyle}
 
\def\gtap{\raisebox{-.4ex}{\rlap{$\sim$}} \raisebox{.4ex}{$>$}} 

\begin{document}

\renewcommand*{\thefootnote}{\fnsymbol{footnote}}

\begin{center}
 {\Large\bf{Unified resolution of the $\bm{R(D)}$ and $\bm{R(D^*)}$ anomalies and the lepton flavor violating decay 
 $\bm{h\to\mu\tau}$} 
} \\[6mm]
 {Debajyoti Choudhury $^1$\footnote{Electronic address: debajyoti.choudhury@gmail.com}, 
 Anirban Kundu $^2$\footnote{Electronic address: akphy@caluniv.ac.in}, 
 Soumitra Nandi $^3$\footnote{Electronic address: soumitra.nandi@gmail.com}
and Sunando Kumar Patra $^3$}\footnote{Electronic address: sunandoraja@gmail.com} \\[3mm]

$^1${\small\em Department of
  Physics and Astrophysics, University of Delhi, Delhi 110007, India}

$^2${\small\em Department of Physics, University of Calcutta,
92 Acharya Prafulla Chandra Road, Kolkata 700009, India}

$^3${\small\em Department of Physics, Indian Institute of Technology, North Guwahati, Guwahati 781039, Assam, India}
 \end{center}


\begin{abstract}

Taking advantage of the fact that the flavor of the neutrino in semileptonic $B$ decays $B\to
D^{(*)}\tau\nu$ is not known, we show how a minimal set of higher-dimensional
lepton flavor violating (LFV) operators can explain the $R(D^{(*)})$ anomalies, and as a 
spin-off, can give rise to the LFV decay of the Higgs boson, $h\to\mu\tau$. We also show how 
none but the minimal set of operators survive the present data.

\end{abstract}



\setcounter{footnote}{0}
\renewcommand*{\thefootnote}{\arabic{footnote}}

\section{Introduction}

The search for signals of lepton flavor violation (LFV) has been a
long and varied quest, for it is believed to not only constitute a
smoking gun for new physics (NP) beyond the Standard Model (SM), but
also shed light on a variety of issues ill-understood within the SM,
such as the origin of flavor on the one-hand and the generation of
non-zero lepton and baryon number in the universe, on the
  other.  While the SM can incorporate LFV, as seen, e.g., in
neutrino oscillations, by the mere inclusion of right-handed neutrino
fields and consequent Dirac masses, the corresponding LFV amplitudes
would be too small to be manifested in processes involving charged
leptons\footnote{It should also be noted that total lepton
    number conservation is an accidental symmetry within the SM, and
    that the inclusion of right-handed neutrino fields would allow for
    unsuppressed Majorana masses as well (unless a global $U(1)_L$ is 
    imposed), thereby further enriching
    the neutrino mass sector. 
    With the Majorana/Dirac masses suffering only logarithmic corrections, 
    ascribing appropriate (small) values to these is technically natural.}. 
Even the proposed upgrades, or new experiments, are expected to improve the 
limits on LFV processes by at most one order of magnitude, except for
$\mu\to 3e$ 
and $\mu$-$e$ conversion \cite{Heeck}.
 Indeed, if decays such as $\mu \to e
\gamma$ or $\tau \to 3 \mu$ are seen in experiments currently in
operation or due to start in the near future, the corresponding
amplitudes would be too large to be supported by such trivial
extensions of the SM.

It is in this context that the recently reported \cite{1502.07400}
hint, from the CMS experiment, of the Higgs boson decay 
$h \to \mu \tau$ is to be
viewed.  If this is not a mere background fluctuation but an actual
signal, one has to entertain the possibility that such LFV decays are
flavor-specific, as neither CMS nor
  ATLAS has seen any LFV in channels like $h\to e\tau$ or $h\to e\mu$
\cite{1607.03561}. This, however, is not unnatural, simply because
such a decay is quite likely to be generated from Yukawa couplings,
and the latter are believed to be typically stronger for the
higher generations, even in extensions of the SM. While the results
from the ATLAS experiment on $h\to\mu\tau$ are more or less consistent
with zero, these too can allow for a nontrivial branching ratio (BR)
for this channel. The measurements have yielded
\cite{1502.07400,1604.07730}
\beq
{\rm BR}(h\to\mu\tau) = 0.84^{+0.39}_{-0.37}\%~~{\rm (CMS)}\,,\ \ 
0.53 \pm 0.51\% ~~{\rm (ATLAS)}\,,
\eeq
so that the 95\% CL upper limits on the BR 
are 1.51\% (CMS) and 1.41\% (ATLAS) respectively.

While the CMS measurement {\em per se.}\ does not call for new
physics right away, it is interesting to juxtapose it against another
long-standing anomaly, albeit in a completely different sector.  
The ratios of the partial widths of $B$ mesons, $R(D)$ and
$R(D^*)$, defined as
\beq
R(D^{(*)}) = 
\frac{{\rm \Gamma}(B\to D^{(*)}\tau\nu)} 
{{\rm \Gamma}(B\to D^{(*)}\ell\nu)}\, ,
\eeq
(with $\ell=e,\mu$) are particularly clean probes of physics beyond
the SM, on account of the cancellation of the leading uncertainties
inherent in individual BR predictions. The values of $R(D)$ and
$R(D^{*})$ as measured by \Babar~\cite{Lees:2013uzd}, when taken
together, exceed SM expectations by more than $3 \sigma$, which
generated interest in the first place.  Furthermore, the Belle
measurements for the same observables lie in between the SM
expectations and the \Babar~measurements and are consistent with both
\cite{Huschle:2015rga}.  Recently, Belle has published their new
result on $R(D^{*})$ \cite{Abdesselam:2016cgx} with $\tau$ decaying
semileptonically, and this agrees with the SM expectations only at the
$1.6 \sigma$ level, while the first measurement by LHCb
\cite{Aaij:2015yra} is also $2.1 \sigma$ above the SM prediction.  Taking all the results 
together, including the correlations, the tension between data and SM is at the level of 
$3.9\sigma$. On
the other hand, the recent results on the measurement of
$\tau$-polarization for the decay $B\to D^*\tau\nu$ in Belle
\cite{Hirose:2016wfn} are consistent with the SM predictions,
albeit with only a large uncertainty.

While the ``anomalies'' in either of $R(D)$ and $R(D^*)$ do not call
for LFV, clearly they seem to be associated with a loss of lepton
universality, and involving the very same fermions as the anomalous
decay. It is therefore conceivable that the individual excesses,
intriguing in their own right but not calling out for a rejection of
the SM, are, together, indicative of some new physics. 
A combined approach to treat both these anomalies together within the scope of a 
particular model may be found in Ref.\ \cite{crivellin-heeck} \footnote{There have been 
numerous attempts to relate the $R(D^{(*)})$ anomaly with some other anomalous observables,
see, {\em e.g.}, Ref.\ \cite{girish}.}. 
At this point, we may refer 
the reader to Refs.\ \cite{Bhattacharya:2016, byakti}, and the references therein, for 
a detailed analysis of the NP operators. In this paper, we investigate this 
more closely, coupled with the LFV Higgs decays. In particular, if anomalous Higgs
interactions are indeed called for, we show that the difference between the
chiral structure of the ensuing four-fermi operators and that of the
SM operator could possibly explain why the experimental
discrepancies are seen only in certain channels.

The generation of such LFV decays
 of the Higgs is relatively simple if
the scalar sector is enlarged, as in a Type-III two-Higgs doublet
model wherein the 125 GeV scalar has a tiny component of the field
responsible for the LFV decays~\cite{crivellin}. A variation is
afforded by scenarios~\cite{dipankar} wherein there are two or more nearly
degenerate scalars with one of them being SM-like and the other(s) having
explicitly LFV couplings. On the other hand, lepton flavor
non-universality can appear in many a guise, whether it be through
Higgs couplings or through gauge couplings in a theory with extended
symmetry or even through the exchange of other non-standard particles
such as superpartners in a supersymmetric extension of the SM, or leptoquarks. 
Hence, rather than adopt any particular scenario, we investigate 
the {\em structure}
of the minimal alteration to the SM that can satisfactorily 
explain the anomalies while remaining consistent with the rest of 
low-energy phenomenology. In other words, we advocate a bottom-up 
approach, starting with an effective theory.

In this paper, we would like to investigate whether both these decays, namely, 
$h\to \mu\tau$ and $B\to D^{(*)}\tau\nu$ can be simultaneously affected by 
a single four-fermion operator, keeping the scalar sector to be completely SM-like 
at the electroweak scale. There are at least two points worth emphasizing, so let us 
note them down here.

\begin{itemize}
 \item If the scalar sector is completely SM-like at all energies,
   {\em i.e.}, if the mass matrix and the Yukawa matrix are
   proportional, there can be no flavor-changing coupling of the Higgs
   boson of the form $h\bar{f}_i f_j$ with $i\not= j$, even at the
   one-loop level. This is in contradiction to what has been claimed
   in, for example, Refs.\ \cite{willey,bird}. The reason is not
   difficult to understand: as soon as one generates an off-diagonal
   Yukawa coupling $h_{ij}$, an analogous term $m_{ij}=vh_{ij}$ is
   also generated in the mass matrix, where $v$ is the vacuum
   expectation value (VEV) for the CP-even neutral component of the SM
   Higgs field $\Phi$. Thus, one needs to redefine the stationary
   basis for the fermions again, and in that new basis, such
   off-diagonal effective Yukawa couplings no longer exist. However,
   there are possible ways out \cite{ellis, harnik}, and we will later
   show, with a toy model, how to achieve this.  In this sense, we
   demonstrate how to generate the LFV decay of the Higgs boson
   without introducing any low-energy extension of the scalar sector.
 
 \item NP has to be there in some form or other at some high scale,
   but if the low-energy sector is SM-like, then any new state can
   exist only at a scale $\Lambda\, \gtap\, {\cal O}(1~{\rm TeV})$, the
   natural scale for NP. It is possible, though, that NP can appear at
   several (well-separated) scales, with the aforementioned $\Lambda$
   being the lowest of them all.
\end{itemize}

Here, we will focus on some possible dimension-6 four-fermion
operators to explain both the anomalies, relating the charged current
operator $b\to c\tau\nu$ with the neutral current operator, that
produces $\tau\mu$ in the final state, through SU(2)$_L$. We will take
advantage of two facts: first, the quark mixing in the right-chiral
sector is essentially unconstrained, and second, the flavor of the
neutrino that comes out in semileptonic $B$ decays is not determined.
While a similar exercise using higher dimensional effective
operators has been performed \cite{cgk}, it was
  restricted only to the $B$-sector observables. The novelty, in our approach, lies
  in that we do not consider any extension of the SM scalar sector,
and the Yukawa couplings remain unchanged.  As we will show, the new
operators that we consider produce an effective $h\mu\tau$ vertex,
which we illustrate with the help of a toy model.  
Showing how experimental constraints already rule out most of the
possible operators, we identify the minimal set of operators necessary to explain the anomalies.

The paper is arranged as follows. In Section II, we will first
describe a toy model to generate flavor-changing Higgs couplings with
lowest dimensional effective operators, and then elaborate our
model. In Section III, we show how it affects the LFV Higgs decay $h\to\mu\tau$, 
and semileptonic $B$ decays are treated in Section IV. 
We summarize and conclude in Section V.


\section{The formalism}

Assuming that the (low-energy) scalar sector is just as in the SM, the
only way to explain a LFV decay of the Higgs boson $h$ (such as the one
under discussion) would be to postulate a term $\left[-y_{ij}
  \bar\ell^i_L \ell^j_R h + {\rm h.c.}  \right]$ (with $i\ne j$), in
the Lagrangian, keeping in abeyance, for the time being, any
discussion of the source of this term. Written in full, the relevant
term is
\beq
-y_{\mu\tau} \left(\bar\mu_L \tau_R + \bar\tau_R \mu_L\right) h  
-y_{\tau\mu} \left(\bar\tau_L \mu_R + \bar\mu_R \tau_L\right) h\, ,
\label{lfv-lag}
\eeq
and the corresponding branching fraction is given by
\beq
{\rm BR}(h\to\mu\tau) = \frac{m_h}{8\pi\Gamma_h} \left( |y_{\tau\mu}|^2 + |
y_{\mu\tau}|^2\right)\,,
\eeq
where $y_{\mu\tau}$ and $y_{\tau\mu}$ are effective Yukawa couplings,
which need not be equal, or even of the same magnitude.  If
$h\to\mu\tau$ (and other possible new decay channels) have only a
small BR, one can assume $\Gamma_h \approx \Gamma_h^{\rm SM} \approx
4.07$ MeV for $m_h \approx 125$ GeV.

If the scalar sector (both the field content and interactions) is
restricted to being exactly as in the SM, clearly, 
terms as in Eq.\ (\ref{lfv-lag}) cannot occur at the tree-level. They may 
appear as quantum corrections though, and the required size clearly does 
not preclude this. However, for even this to work, either the field 
content of the theory has to be enlarged or non-renormalizable interactions 
introduced or both.

\subsection{Flavor-changing Higgs couplings: A toy model}

As was discussed earlier, one cannot simply postulate such an
off-diagonal coupling for the Yukawa and the mass matrices often turn
out to be proportional to each other (not only at the tree level, but
to any given order in perturbation theory).  To circumvent this
argument, let us consider a toy model.  Suppose the
Lagrangian contains dimension-5 terms like
\beq
\frac{1}{\L}\left[ a_t \bar t_R {Q}_L \tilde\Phi X + 
a_l \bar \tau_R {L}_L \Phi X^* \right] + {\rm H.c.}
\eeq
where $\Phi$ is the SM doublet ($\tilde\Phi = i\sigma_2 \Phi^*$), and
$X$ is a complex SU(2)$_L$ triplet with hypercharge $Y=2$.  We will
assume that the mass-squared term for $X$ is positive and ${\cal
  O}({\rm TeV}^2)$. Consequently, the components of $X$ receive no
vacuum expectation value, thereby trivially satisfying the constraints
from the $\rho$-parameter. A further consequence is that they are
almost degenerate in mass, which allows the scenario to evade the
remaining constraints from electroweak precision observables.  $\L$
above is a cutoff scale, with $\L \gg m_X$ so as to validate the
effective Lagrangian approach.

Written in full, with $X=(x^{++}, x^+, x_0)$, the relevant terms look like 
\beq
{\cal L} \supset \frac{1}{\sqrt{3}\L}\left[a_t \left(
\bar{t}_R t_L \phi^{0*} x_0 - \frac{1}{\sqrt{2}}\bar t_R{b}_L \phi^{0*} x^+\right) + 
a_l\left(\bar \tau_R \mu_L \phi^0 x_0^* - \frac{1}{\sqrt{2}}\bar\tau_R \nu_{\mu L} \phi^0 x^- \right)
\right] + H.c. \,.
\label{lag-dim5}
\eeq
Integrating out the $X$ fields yields a dimension-8 term in the
Lagrangian of the form
\beq
\frac{- a_t a_l} {3\L^2 m_X^2} |\phi^0|^2 \; \bar{t}_L t_R\;  \bar\mu_L \tau_R + h.c.,
\eeq
valid at scales well below $m_X$. Here, analogous terms involving the
putative Goldstones have been suppressed.  On the breaking of the
electroweak symmetry, one may write $\phi^0 = (h+v)/\sqrt{2}$, with
$h$ being the physical Higgs field.  This yields not only a four-Fermi
term of the form
\beq
{\cal L}_{\rm 4fer} = \frac{a_t a_l}{6\L^2}\, \frac{v^2}{m_X^2}\, \left(
\bar{t}_L t_R\right) \left(\bar{\tau}_R \mu_L\right) + {\rm H.c.},
  \label{four_fermi_toy}
\eeq
 but also couplings of the same set of fields 
with both a single higgs and a
pair of higgses, or, in other words, a five-field and 
a six-field vertex each.  Of immediate concern are the first two of these
terms. Clearly the $(2vh) \, \bar{t}_L t_R \bar \mu_L \tau_R$
term, on contracting the top-fields, would lead to an effective LFV
coupling $h \bar\mu_L \tau_R$. Similarly, the 
term in Eq.\ (\ref{four_fermi_toy}) 
would contribute to an off-diagonal mass term
connecting the muon and the tau.  Importantly, these one loop
contributions to the Yukawa and the mass matrices bear a relation
different from the tree-level terms, viz.  $\delta y_{\mu\tau} = 2 \,
\delta m_{\mu\tau} / v$. The extra factor of 2 destroys the overall
proportionality of the Yukawa and the mass matrices, thereby allowing
for a LFV Higgs coupling when the fermions are rotated into the
stationary basis.

The evaluation of the loop contributions is quite straightforward.
While they are, formally, quadratically divergent, it needs to be
realized that the effective theory under consideration has a natural
cutoff at $m_X$.  The leading term, apart from the overall coupling,
is thus $-4N_c m_X^2 m_t/16\pi^2$, where the minus sign comes from the
fermion loop and $m_t$ from the chirality flip.  Thus, the effective
LFV Yukawa coupling is given by
\beq
\frac12 \times \frac{a_t \, a_l\, v}{3\, \L^2} \, \frac{N_c}{4\pi^2}\, 
                m_t \, \bar\mu_L \tau_R h\,.
\eeq 
The factor of half  needs explaining. As mentioned above,
the term proportional to $v^2$ generates an off-diagonal term in the
mass matrix and, consequently, an extra rotation is needed to get back
to the new mass basis. This absorbs half of the effect (which is why a
coupling proportional to $(h+v)$ cannot lead to flavor-changing Yukawa
couplings), leaving us with the remaining half.

It should be noted that much the same low-energy phenomenology could
have been obtained, had we started with an $Y=0$ triplet instead, with the
Lagrangian now being
\[
\frac{1}{\L}\left[ a_t \bar t_R {Q}_L \Phi X + 
a_l \bar \tau_R {L}_L \tilde\Phi X^* \right] + {\rm H.c.} \ .
\]
Similarly, had we started with a scalar leptoquark field, 
coupling to both a $t$-$\tau$ and a $t$-$\mu$ current, the ensuing 
effective Lagrangian, on Fierz-rearrangement, would yield terms 
analogous to those above, but with (axial-)vector couplings instead.

\subsection{The minimal operator basis}

Having argued that it is indeed possible to generate flavor-changing
Higgs couplings (for a theory with a single scalar doublet)
within the stationary basis, and that this may be
achieved quite naturally within the paradigm of an effective theory,
we now turn to the other anomalies at hand, namely $R(D^{(*)})$.  To
this end, we augment the SM by postulating at most a couple of
effective dimension-6 operators obeying the full symmetry of the SM.
These operators will be shown to generate an effective $h\mu\tau$
vertex, by a mechanism similar to that outlined above, which is of the
right magnitude.  While a similar approach was adopted in
Ref.\cite{pilaftsis} to explain $h \to \mu \tau$ alone, 
we go much beyond and relate the operators to
the anomalies in $R(D)$ and $R(D^*)$.

Following Refs.\ \cite{sakaki,srimoy}, let us consider an effective
charged-current Hamiltonian of the form
\beq\label{hamil}
{\cal H}_{\rm eff} = \frac{4G_F}{\sqrt{2}} V_{cb} \left[O_{\rm SM} + C_{S_1} O_{S_1} + 
C_{S_2} O_{S_2} + C_T O_T\right]\,,
\eeq
where
\beq
\barr{rcl}
\dis O_{\rm SM} &=& \dis 
    (\bar c_L\gamma^\lambda b_L) (\bar \tau_L \gamma_\lambda \nu_{\tau L})\,,
\\[1.5ex]
\dis O_{S_1} &=& \dis (\bar c_L b_R) (\bar \tau_R \nu_{\mu L})\,,\\[1.5ex]
\dis O_{S_2} &=& \dis (\bar c_R b_L) (\bar \tau_R \nu_{\mu L})\,,\\[1.5ex]
\dis O_T &=& \dis (\bar c_R \sigma^{\mu\lambda} b_L) (\bar \tau_R \sigma_{\mu\lambda} \nu_{\mu L})\, ,
\earr
   \label{operator_list}
\eeq
and the fermion fields are weak-eigenstates, as befits operators in an
effective theory defined above the electroweak scale.  While $O_{S_1}$
and $O_{S_2}$ might result from the mechanism discussed in the
previous subsection (albeit with different fermionic fields), the
generation of $O_T$ is more non-trivial, and the ultraviolet
completion of the same would, typically, require the introduction of
exotic fields\footnote{It should be noted, though, that such a
  rendition would require the simultaneous introduction of other
  operators as well.}, such as a doublet scalar leptoquark with a
hypercharge of $\frac76$.  Note that this set is not exactly identical to
that given in Ref.~\cite{srimoy}. For one, the new operators contain
$\nu_\mu$ instead of $\nu_\tau$.  With the neutrinos in a decay being
unidentified, this does not affect the analysis of $R(D)$ and $R(D^*)$
except for the fact that, now, no interference between the SM operator
$O_{\rm SM}$ and the new operators would exist. Furthermore, we have
dropped some operators, involving (axial-)vector currents, as they (to
be demonstrated shortly) not only do not lead to $h \to \mu \tau$,
but, in addition, cause disagreements with other observables. Later on, 
we will show that $C_{S_1}$ should be of the order of unity to produce a good
fit with the data, and it is almost trivial to show that this leads to an unacceptably 
large contribution to the decay $B_s\to\mu\tau$, which is yet to be observed. Thus,
even the operator $O_{S_1}$ falls out of favor, but we will keep this in our analysis 
for the time being.

The origin of the specific set of operators is, of course,
uncertain. Given that the family number is conserved, it is quite
conceivable, for example, that these arise on account of flavor
dynamics. We do not, however, attempt to answer such questions, but
only offer the argument that this leads us to the minimal set of new
operators required to explain the data.  To further reduce the number
of free parameters, we shall consider an additional simplification and
consider two reduced sets, namely
\begin{itemize}
\item Model 1: $C_{S_1}$, $C_{S_2}\not = 0$, $C_T=0$; 
\item Model 2: $C_{S_2}$, $C_T\not=0$, $C_{S_1}=0$. 
\end{itemize}
In other words, only two new Wilson coefficients are introduced in
each case.  

The new operators also imply the existence of their SU(2) conjugates, 
with identical Wilson coefficients, 
namely
\beq
\barr{rcl}
\dis O'_{S_1} &=& \dis (\bar s_L b_R) (\bar \tau_R \mu_{L})\,,\\[1.5ex]
\dis O'_{S_2} &=&\dis  (\bar c_R t_L) (\bar \tau_R \mu_{L})\,,\\[1.5ex]
\dis O'_T &=& \dis (\bar c_R \sigma^{\mu\lambda} t_L) (\bar \tau_R \sigma_{\mu\lambda} \mu_{L})\,.
\earr
     \label{rotated_op}
\eeq
This, immediately, puts into perspective our earlier assertion about
$O_{S_1}$ being highly constrained, for $O'_{S_1}$ would readily
generate semileptonic LFV decays like $B\to K^{(*)}\tau\mu$ and the
purely leptonic decay $B_s\to\tau\mu$. In fact, if the corresponding
Wilson coefficient $C_{S_1}$ is of order unity, the BR of
$B_s\to\tau\mu$ becomes so large ($\sim {\cal O}(0.1)$) that it should
certainly have been observed. Thus, unless $C_{S_1}$ is of the order
of at least $10^{-3}$, it is hard, but not entirely impossible, to
entertain $O_{S_1}$ (and hence Model I as mentioned before) as a
possible candidate for the minimal set of operators.

Having been written in terms of the weak-interaction eigenstates, the
operators need to be re-expressed in terms of the stationary states
(i.e., the mass eigenstates). With the fermion mass-matrices being
diagonalized through a bi-unitary transformation, we have, in
principle, as many as four $3 \times 3$ unitary matrices ($U_{L,R}$,
$D_{L,R}$) in play, one each for the (left-) right-handed (up-)
down-quarks. Thanks to the right-handed fields being $SU(2)_L$
singlets and universality of the gauge-structure across generations,
within the SM, two of these matrices ($U_R$ and $D_R$) play no dynamic
role, and only the combination $U_L^\dagger D_L$ is manifested
physically (as the Cabibbo-Kobayashi-Maskawa matrix). In the presence
of these new operators, this would no longer be the case. In
particular, both of $U_R$ and $D_R$ would now play a nontrivial
role. Once again, rather than consider the most general case, we
simplify the analysis by retaining only the most important term,
namely
\beq
c_R = \cos\alpha\, c'_R + \sin\alpha\, t'_R\,, \ \ t_R = -\sin\alpha\, c'_R + \cos\alpha\, t'_R\,,
\eeq
where the primed fields are in the mass basis. This immediately leads to
\beq
\barr{rcl}
\dis O_{S_2} &=& \dis \cos\alpha \; (\bar c'_R b_L) (\bar \tau_R \nu_{\mu L}) + \cdots\,,\\[1.5ex]
\dis O_T &=& \dis \cos\alpha \; (\bar c'_R \sigma^{\mu\lambda} b_L) (\bar \tau_R \sigma_{\mu\lambda} \nu_{\mu L}) + \cdots\,,\\[1.5ex]
\dis  O'_{S_2} &=& \dis \sin\alpha \; (\bar t'_R t_L) (\bar \tau_R \mu_{L}) + \cdots \,, \\[1.5ex] 
\dis O'_T &=& \dis \sin\alpha \; (\bar t'_R \sigma^{\mu\lambda} t_L) (\bar \tau_R \sigma_{\mu\lambda} \mu_{L}) + \cdots\,.
\earr
\label{all-op}
\eeq
The left-chiral quark fields are also rotated to the mass basis as per
the Cabibbo-Kobayashi-Maskawa paradigm. These rotations have
  important physical consequences. For example, even if the mixing
is confined to the down quark sector alone,
$O'_{S_1}$, after field rotation, can lead to $\Upsilon\to\mu\tau$,
which, within the SM, is highly suppressed compared to the electromagnetic
decay $\Upsilon\to\ell^+\ell^-$. This particular mode, though,  
is not very restrictive once the aforementioned constraints from 
$B_s \to \tau \mu$ are satisfied. Similarly, if the mixing is for the up-type
quarks, $O'_{S_2}$ and $O'_T$ can lead to LFV charmonium decays, which
are also yet to be observed.  While eq.\ (\ref{all-op}) lists all the
operators relevant for our study, it is instructive, at this stage, to
examine the ramifications thereof. Clearly, engendering the
flavor-changing Yukawa coupling $h \bar \mu \tau$ by Wick-contracting
the top-fields is possible only for $O'_{S_2}$. Thus, only this
operator (and its sibling, $O_{S_2}$) are relevant for this aspect. On
the other hand, $O^{(')}_{S_1}$ and $O^{(')}_{T}$ appear at the same
order in the effective theory and, like $O^{(')}_{S_2}$, can
contribute to both $R(D)$ and $R(D^\ast)$.  Thus, the inclusion of at least
two operators is necessary to maintain agreement for these decays.

Before we end this section, we would like to point out out that, in
obtaining the operators in Eq.\ (\ref{rotated_op}) from those in
Eq.\ (\ref{operator_list}) through basis transformations, we would
also generate many other operators, designated by the ellipses in
Eq.\ (\ref{rotated_op}). These would have their own consequences, such
as the FCNC top decay $t\to c\mu\tau$.  We have checked that, for the
sizes of the Wilson coefficients ($C_{S_2}$ 
accompanied by one of $C_{S_1}$ and
$C_T$) that we would need, such effects are negligible.


\section{LFV decays of the Higgs}

\begin{figure}[!h]
\begin{center}
\includegraphics[width=10cm]{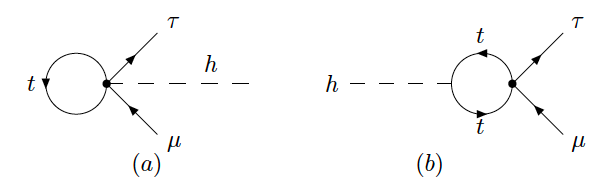} 
\caption{Typical contributions
 to the decay $h \to \mu^+ \tau^-$ initiated
  by the new operators.  Diagrams for the conjugate process would
  be analogous.}
  \label{fig:Feynman1}
\end{center}
\end{figure}

The presence of an operator such as $(\bar f \Gamma_a f) \, (\bar \tau
\Gamma_a \mu)$, where $f$ is a SM fermion and $\Gamma_a$ a Dirac
matrix, denotes the violation of both $N_\tau$ and $N_\mu$ while
preserving their difference. Clearly, this can result in $h \to
\mu\tau$, at least at the loop-level. Fig.\ \ref{fig:Feynman1} shows
two typical diagrams, in the context of the toy model
discussed before, that contributes to such a process.

It is easy to see that $O_T$ cannot contribute to this amplitude, for,
to obtain a $h\mu\tau$ vertex, we would need to contract the leptonic
current with two external momenta which, of course, is not possible.
For (axial-)vector operators (not listed in Eq.\ (\ref{operator_list})),
on the other hand, only one such contraction is needed and,
consequently, the amplitude is proportional to the lepton
mass. Furthermore, the very structure of the operator ensures that the
loop integral is logarithmically divergent and scales only as 
$m_X^{-2} \, \ln(m_h^2 / m_X^2)$ at the most. While this suppression is
not necessarily an overwhelming one (provided $m_X$ is not too
large), it should be realized that corresponding diagrams exist where
the Higgs field is replaced by the $Z$. The latter would lead to an
unsuppressed contribution \cite{Choudhury:2013jta} to the decay $Z \to
\tau \mu$, well beyond the experimental limits, unless the Wilson
coefficient for the four-fermion interaction is suppressed
enough. This, though, would imply that the operator has a negligibly
small effect in Higgs decays.

This leaves us with the (pseudo-)scalar operators $O_{S_1}$ and $O_{S_2}$.
Let us concentrate on the latter, and take our toy model as a concrete example. 
This  gives 
\beq
\frac{4 G_F}{\sqrt{2}}V_{cb} C_{S_2}\, \sin\alpha =
\frac{a_t a_l}{6\Lambda^2} \, \frac{v^2}{m_X^2}\,,
\eeq
and hence, the first diagram of Fig.\ref{fig:Feynman1} yields 
\beq
y_{\mu\tau} = \frac{G_F}{\sqrt{2}\pi^2}\, m_X^2 V_{cb} N_c \frac{m_t}{v}\, 
C_{S_2}\, \sin\alpha \approx 0.076 \left( \frac{m_X}{1~{\rm TeV}}\right)^2 
C_{S_2}\, \sin\alpha\,,
\eeq
where $N_c=3(1+\alpha_s/\pi)\approx 3.11$ is the effective number of
colors, and $h_t\approx 1$ is the top quark Yukawa coupling.  We have
also used $m_t=175$ GeV, $G_F = 1.16637\times 10^{-5}$ GeV$^{-2}$, and
$|V_{cb}|= (41.1\pm 1.3)\times 10^{-3}$. The contribution of the
  second diagram of Fig. \ref{fig:Feynman1} is further suppressed by a
  factor of $\sim v/m_X$.
This gives 
\beq
{\rm BR}(h\to\mu\tau) \approx 7.1  \left( \frac{m_X}{1~{\rm TeV}}\right)^4 \left[
C_{S_2}\sin\alpha\right]^2 < 0.014 \Rightarrow C_{S_2}\, \sin\alpha < 4.4\times 10^{-2}\left( 
\frac{1~{\rm TeV}}{m_X}\right)^2\,.
\label{lfvlim1}
\eeq 
Thus, if $|C_{S_2}|$ is of order unity, one needs a small mixing in the $t_R$-$c_R$ sector,
namely, $\tan\alpha \sim 10^{-2}$, to explain the LFV Higgs decay. 
Note that while the estimation has been done
for a particular toy model, the essence is model-independent.

\section{The $B$-decay anomalies}

In terms of the differential distributions $d\Gamma/d q^2$ for 
the decay $B\to X\ell\nu$, where $q_\mu \equiv (p_B - p_X)_\mu$ 
is the momentum transfer, the ratios $R(D)$ and $R(D^*)$ are defined as
\beq
 R(D^{(*)}) = \dis
\left[\int^{q^2_{max}}_{m^2_{\tau}} \frac{d\Gamma\left(\overline{B} \rightarrow D^{(*)} 
 \tau \overline{\nu}_{\tau}\right)}{d q^2} d q^2\right]  \;
\left[\int^{q^2_{max}}_{m^2_{\ell}} 
 \frac{d\Gamma\left(\overline{B} \rightarrow D^{(*)} \ell \overline{\nu}_{\ell}\right)}{d q^2} d q^2\right]^{-1}
 \label{Rth}
\eeq 
with $q^2_{max}= (m_B - m_{D^{(*)}})^2$, and $\ell=e$ or $\mu$. In
each case, both isospin channels are taken into account.  Using the
effective Hamiltonian in Eq.\ (\ref{hamil}), the expressions for these
distributions are given as
\begin{align}
 \n \frac{d\Gamma \left(\overline{B} \rightarrow D \tau \overline{\nu}_{\tau}\right)}{d q^2} 
 &= 
 \frac{G^2_F \left|V_{cb}\right|^2}{192 \pi^3 m^3_B} q^2 \sqrt{\lambda_D(q^2)} 
 \left(1 - \frac{m^2_{\tau}}{q^2}\right)^2 \times \left\{ 
 \left[ \left(1 + \frac{m^2_{\tau}}{2 q^2}\right) {H^{s}_{V,0}}^2 + 
 \frac{3}{2} \frac{m^2_{\tau}}{q^2} {H^{s}_{V,t}}^2\right] \right.\\
 &\left. +\frac{3}{2} \left|C_{S_1} + C_{S_2}\right|^2 {H^{s}_S}^2 + 
 8 \left|C_T \right|^2 \left(1 + \frac{2 m^2_{\tau}}{q^2}\right) {H^{s}_T}^2 
 \right\}\,,
 \label{dgambd}
\end{align}
and
\begin{align}
 \n \frac{d\Gamma \left(\overline{B} \rightarrow D^* \tau \overline{\nu}_{\tau}\right)}{d q^2} 
 &= \frac{G^2_F \left|V_{cb}\right|^2}{192 \pi^3 m^3_B} q^2 \sqrt{\lambda_{D^*}(q^2)} 
 \left(1 - \frac{m^2_{\tau}}{q^2}\right)^2 \\
 \n &\times \left\{ 
 \left[\left(1 + \frac{m^2_{\tau}}{2 q^2}\right) 
 \left(H^2_{V,+} + H^2_{V,-} + H^2_{V,0}\right) + \frac{3}{2} \frac{m^2_{\tau}}{q^2} H^{2}_{V,t}\right] \right. \\
  & \left.
 + \frac{3}{2} \left|C_{S_1} - C_{S_2}\right|^2 H^2_S + 8 \left|C_T\right|^2 \left(1 + \frac{2 m^2_{\tau}}{q^2}\right) \left(H^2_{T,+} + H^2_{T,-} + H^2_{T,0}\right) 
 \right\}\, ,
 \label{dgambdst}
\end{align}
with $\lambda_X (q^2) \equiv m_B^4 + m_X^4 + q^4 - 2 m_B^2 m_X^2 - 2
m_B^2 q^2 - 2 m_X^2 q^2$. Here, $H_i$s are the respective form factors
as defined within the Heavy Quark Effective Theory
\cite{Caprini:1997mu}, and we use the values determined by the Heavy
Flavor Averaging Group (HFAG) \cite{Amhis:2014hma}. For more details,
we refer the reader to Ref.\ \cite{sakaki}.  While the results for the
lighter leptons are obtained by substituting $m_\tau \to m_\ell
\approx 0$, putting all the $C_i$s equal to zero would yield the SM
results.


\subsection{$R(D)$ and $R(D^*)$}

\begin{figure}[!hbt]
\centering
 \includegraphics[height=6cm]{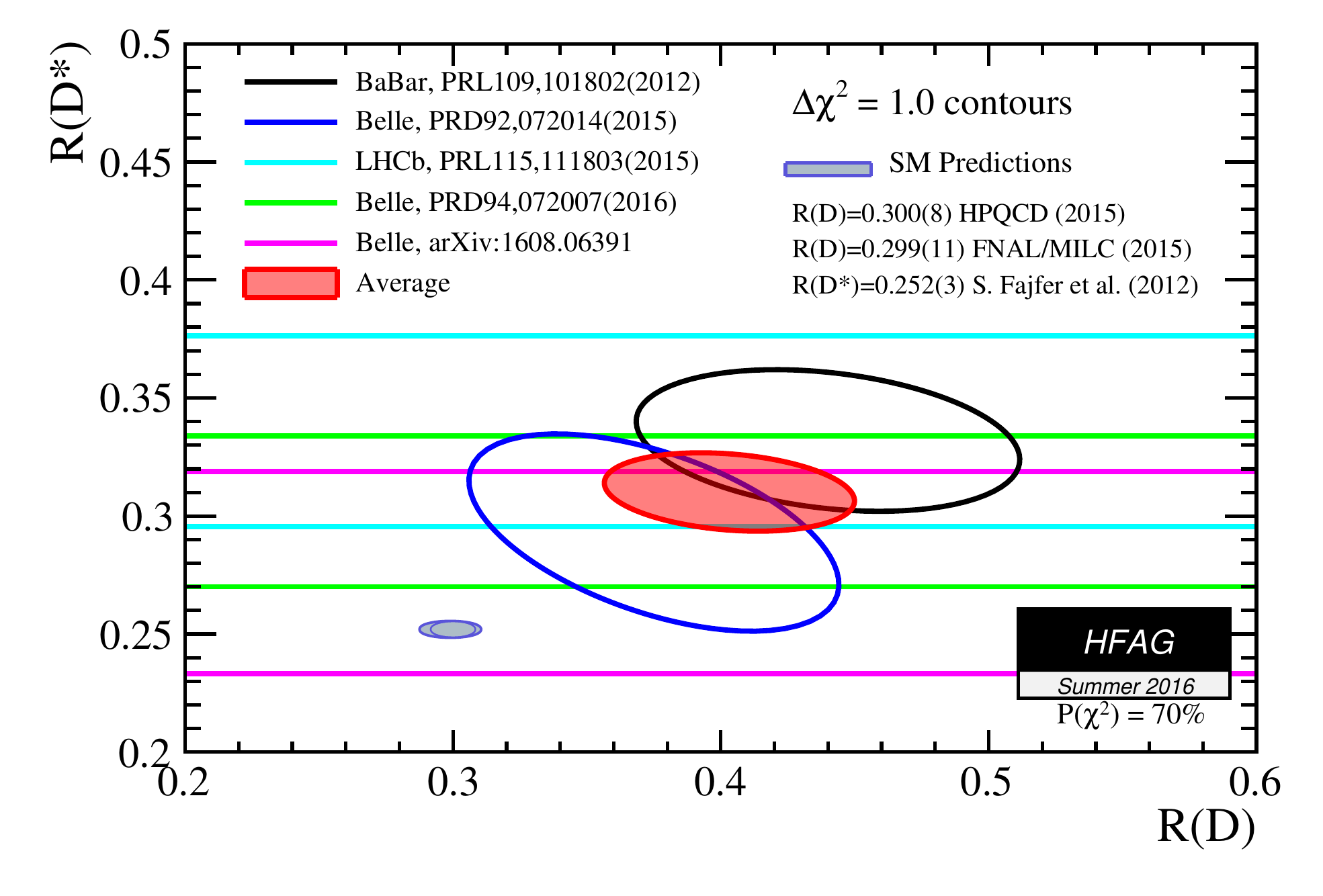}
 \caption{Current experimental status in the measurements of $R(D)$ and $R(D^*)$ \cite{hfag}.}
 \label{fig_hfag}
\end{figure}

Let us first focus on $R(D)$ and $R(D^*)$.  Several experiments have
measured these ratios, and the current status is summarized in
Fig.\ \ref{fig_hfag} as well as in Table \ref{tab:RDRDst}. However, while 
Table \ref{tab:RDRDst} includes the latest Belle result \cite{Hirose:2016wfn} 
on $R(D^*)$, Fig.\ \ref{fig_hfag} takes into account only the Belle update till August 2016. 
Though the change is quite small and can easily be neglected, we 
have used the updated result \cite{Hirose:2016wfn} in our analysis.

\begin{table}[!hbt]
\begin{center}
\renewcommand{\arraystretch}{1.2}
\begin{tabular}{lll}
\hline
\hline
        & ${R}(D)$                      & ${R}(D^*)$ \\
\hline
\noalign{\vskip1pt}
SM  prediction   & $0.300 \pm 0.008$ \cite{Na:2015kha} &  $0.252 \pm 0.003$ \cite{Kamenik:2008tj} \\
\Babar~  (Isospin constrained)  & $0.440 \pm 0.058 \pm 0.042 $ & $0.332 \pm 0.024 \pm 0.018 $ \cite{Lees:2013uzd}\\
Belle (2015)  & $0.375 \pm 0.064 \pm 0.026 $ & $0.293 \pm 0.038 \pm 0.015 $ \cite{Huschle:2015rga}\\
Belle (2016)  & -& $0.302 \pm 0.030 \pm 0.011 $ \cite{Abdesselam:2016cgx}\\
Belle (2016, Full Dataset)  & -& $0.270 \pm 0.035 ~^{+ 0.028}_{-0.025} $ \cite{Hirose:2016wfn}\\
LHCb    & - & $0.336 \pm 0.027 \pm 0.030 $ \cite{Aaij:2015yra}\\
\hline
\hline
\end{tabular}
\caption{The SM predictions for and the data on $R(D)$ and $R(D^*)$. 
While \Babar~considers both charged and neutral $B$ decay channels, LHCb and Belle results,
as quoted here, are based 
only on the analysis of neutral $B$ modes.} 
\label{tab:RDRDst}
\end{center}
\end{table}

While the two scenarios ($C_{S_1} = 0$ vs. $C_T = 0$) are identical as
far as $h\to\mu\tau$ is concerned, their effects are quite markedly
different on $R(D^{(*)})$. We perform a $\chi^2$ goodness-of-fit analysis to fit the
new physics Wilson coefficients through their effects as summarized in
Eqs.\ \ref{dgambd} and \ref{dgambdst}. In our analysis, we use the
$q^2$-integrated data on $R(D)$ and $R(D^*)$, given in Tables
\ref{tab:RDRDst} and \ref{tab:addinput} for different isospin channels
({\em i.e.}, both $B^+$ and $B^0$ decays) with appropriate
correlations wherever the data is available.  However, we have not
used the isospin-constrained data measured by \Babar~(given in Table
\ref{tab:RDRDst}) as an input in our analysis as those are not
independent data-points.  Our analysis involves 11 data-points: 4 from
Ref.\ \cite{Lees:2013uzd}, 2 from Ref.\ \cite{Huschle:2015rga}, 2 from
Ref.\ \cite{Bozek:2010xy}, and 1 each from
Refs.\ \cite{Abdesselam:2016cgx},
\cite{Aaij:2015yra}, and \cite{Hirose:2016wfn}. 
Ref.\ \cite{Bozek:2010xy} supplies the data in
the form of branching fractions. We have converted them to
$R(D^{(*)})$ by normalizing them with ${\rm BR}(B \to D^{(*)} \ell
\nu)$ \cite{b2dlnu} while propagating the errors.


\begin{table}[!hbt]
\begin{center}
\renewcommand{\arraystretch}{1.2}
\begin{tabular}{lll}
\hline
\hline
 Experiment & Channel     & $R(D^{(*)})$ \\
\hline
\noalign{\vskip1pt}
  &  $B^- \to D^0 \tau^- \bar{\nu_{\tau}}$ &$0.429 \pm 0.082 \pm 0.052 $ \\
\Babar~\cite{Lees:2013uzd} &  $\bar{B^0} \to D^+ \tau^- \bar{\nu_{\tau}}$ &  $0.469\pm 0.084\pm 0.053$ \\
      &$B^- \to {D^*}^0 \tau^- \bar{\nu_{\tau}}$ & $0.322 \pm 0.032 \pm 0.022$ \\
      & $\bar{B^0} \to {D^*}^+ \tau^- \bar{\nu_{\tau}}$ &  $0.355\pm 0.039\pm 0.021$ \\
\hline       
 Belle \cite{Bozek:2010xy}  & $B^- \to D^0 \tau^- \bar{\nu_{\tau}}$ &$0.339 \pm 0.112 $  \\
         & $B^- \to {D^*}^0 \tau^- \bar{\nu_{\tau}}$ & $0.372 \pm 0.071$ \\  
\hline
\end{tabular}
\caption{The measured values of $R(D^*)$ in different isospin channels. 
Only Belle 2010 and not the later Belle papers gives the isospin break-up.}
\label{tab:addinput}
\end{center}
\end{table}
%
 \begin{figure}[htbp]
  \begin{center} 
\includegraphics[height=6cm]{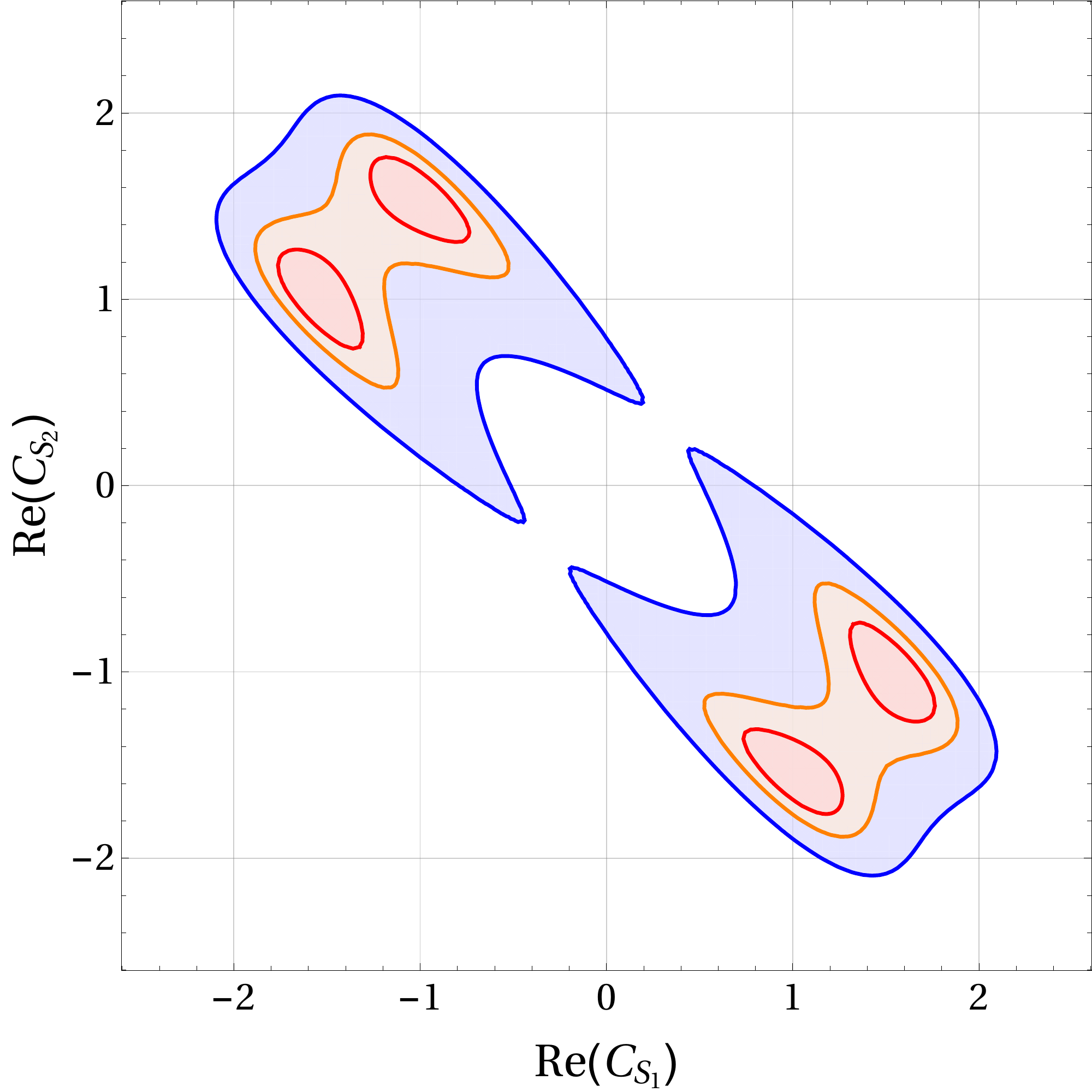} \hspace*{1cm}
\includegraphics[height=6cm]{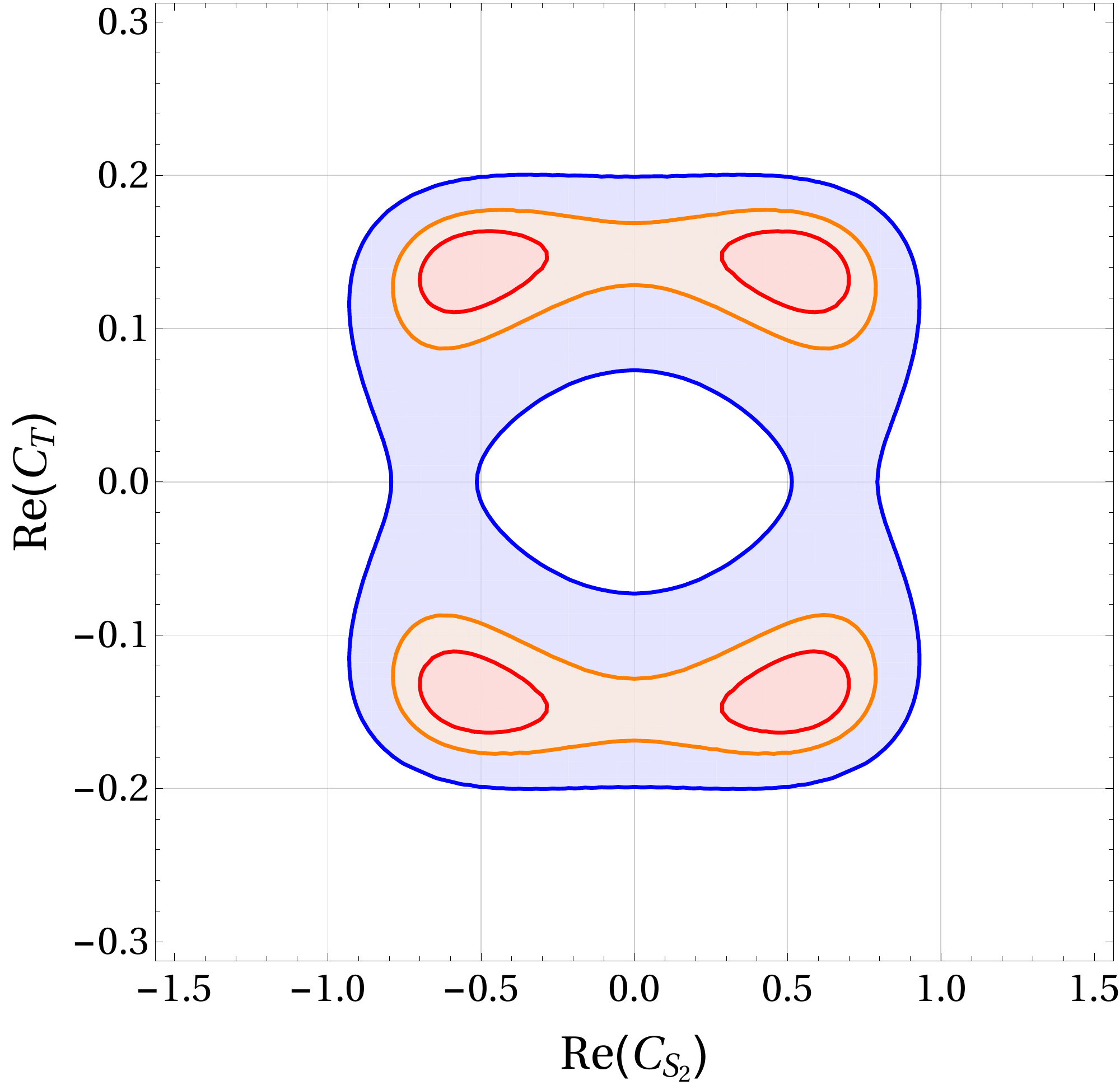}
\caption{The $\chi^2$ contours for Model 1 (left) and Model 2 (right). The $1\sigma$ 
(68.27\%), $2\sigma$ (95.45\%), 
and $4\sigma$ (99.99\%) confidence levels are shown by red, orange, and blue lines 
respectively.}
\label{chisq}
 \end{center}
\end{figure}

An important point to note is that the expressions depend only on
$|C_{S_1}|$ and $|C_{S_2}|$ (or $|C_T|$ and $|C_{S_2}|$) and hence
there is a fourfold ambiguity on the position of the minimum. 
This is best understood from the $\chi^2$ contours shown in Fig.\ \ref{chisq}.
For example, the best fit points are
\begin{align}
{\rm Model~1 :}& \nonumber\\
C_{S_1}\cos\alpha &= \pm(1.55\pm 0.11)\,, \ \ C_{S_2}\cos\alpha = 
-{\rm sgn}(C_{S_1}\cos\alpha) (1.01\pm 0.12)\,,\nonumber\\
{\rm or}~~~ C_{S_1}\cos\alpha &= \pm(1.01\pm 0.12)\,, \ \ C_{S_2}\cos\alpha = 
-{\rm sgn}(C_{S_1}\cos\alpha) (1.55\pm 0.11)\,,\nonumber\\
&{\rm Correlation ~coefficient} = - 0.71 \\
{\rm Model~2 :}& \nonumber\\
|C_{S_2}\cos\alpha| &= 0.53\pm 0.09\,, \ \ ~~~~ |C_T\cos\alpha| = 0.14\pm 0.01\,,\nonumber\\
&{\rm Correlation ~coefficient} = - 0.29 
\label{fitpara}
\end{align}
with almost identical $\chi^2/{\rm d.o.f} \approx 4.50/9$, whereas the SM has $\chi^2 = 33.05$. 
From the smallness of $\alpha$, it is clear that Model 1, with the operator $O_{S_1}$, 
is almost ruled out from the non-observation of $B_s\to\mu\tau$. 

\begin{figure}[htbp]
  \begin{center} 
\includegraphics[height=6cm]{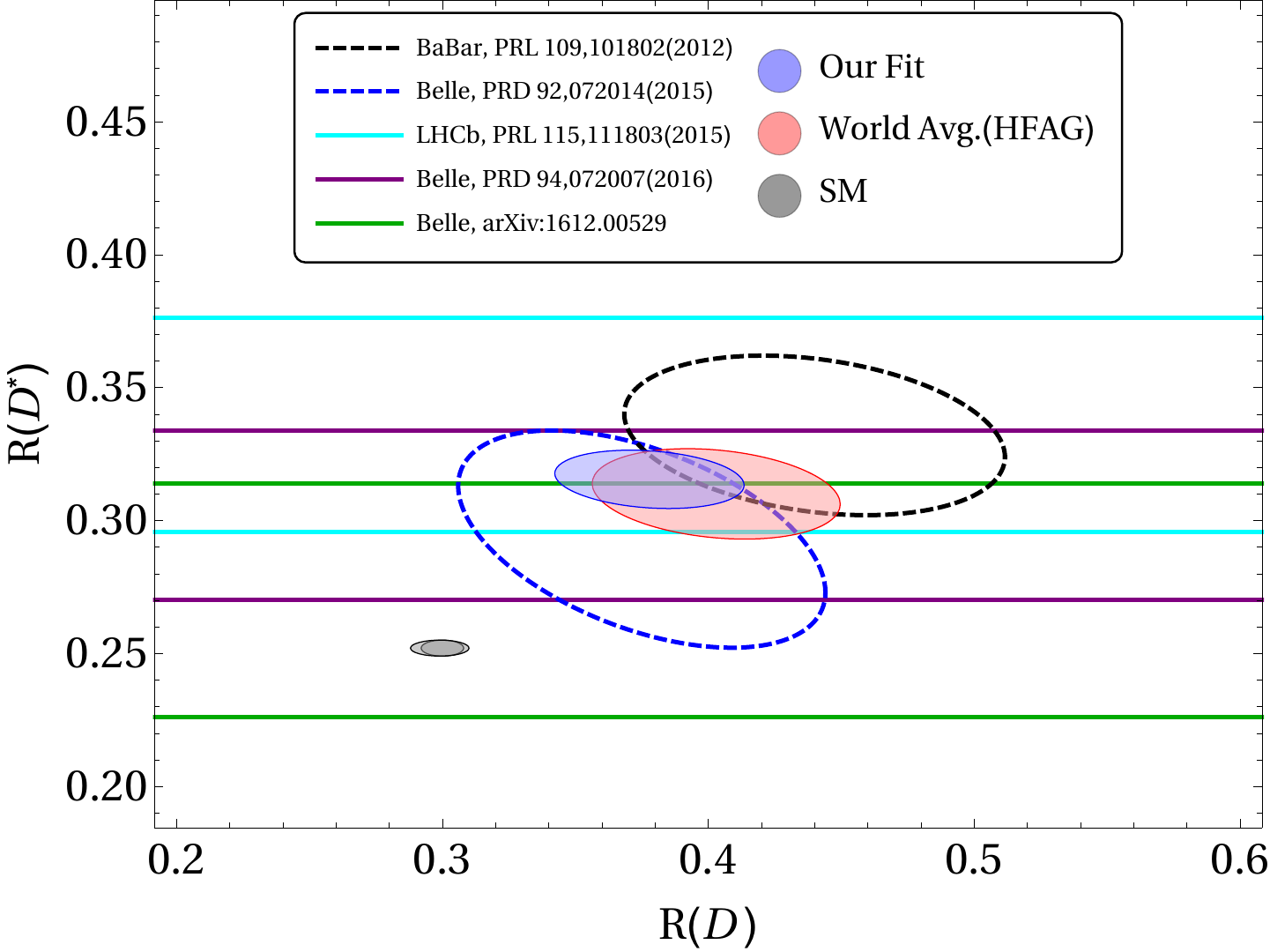}
\caption{The $1\sigma$ contour in the $R(D)$-$R(D^*)$ plane with the best fit points 
for Model 2. The current experimental results and the world average are also shown for comparison.}
\label{fig:plot}
 \end{center}
\end{figure}

For the best fit points, the values of $R(D)$ and $R(D^*)$ are given in Table \ref{tab:predictions}.
We also show, in Fig.\ \ref{fig:plot}, how the $1\sigma$ contours in the $C_{S_2}$-$C_T$ plane translate 
to the $R(D)$-$R(D^*)$ plane. The plot is for Model 2, but it would have been the same for Model 1 
if it were not disfavored, as the goodness-of-fit is the same in both cases. While the operator $O_{S_2}$ 
can lead to the chirally unsuppressed decay through weak annihilation $B_c\to\tau\nu$, whose 
partial width is bounded from the lifetime of the $B_c$ meson \cite{grinstein}, it is easy to check that 
the Wilson coefficient $C_{S_2}$ is not so large as to put that bound in jeopardy. 

\begin{table}[!hbt]
\begin{center}
\renewcommand{\arraystretch}{1.2}
\begin{tabular}{llll}
\hline
\hline
 Decay & Model  & $R(D)$ & $R(D^*)$\\
\hline
\noalign{\vskip1pt}
{\rm From} $B^+$ &  1  &  $0.419\pm 0.072$ & $0.317\pm 0.008$  \\
          & 2 & $0.419\pm 0.040$  & $0.317\pm 0.011$  \\
\hline          
 {\rm From}~$B^0$  & 1 & $0.377\pm 0.064$ & $0.316\pm 0.008$   \\
      &2 & $0.377\pm 0.036$ & $0.316\pm 0.011$ \\
\hline
\end{tabular}
\caption{New physics model predictions of $R(D^{(\ast)})$ with the fitted Wilson coefficients as given in
Eq.\ (\ref{fitpara}).}
\label{tab:predictions}
\end{center}
\end{table}

\section{Conclusions}

In this paper, we have tried to explain, with the introduction of a
minimal set of operators, two apparently uncorrelated anomalies. The
first one is that of the normalized $B\to D^{(*)}\tau\nu$ decay
widths, denoted as $R(D)$ and $R(D^*)$, for which almost all the
experiments find a nontrivial pull from the SM expectations. The
second one is the hint of the LFV decay $h\to\mu\tau$ as seen by the
CMS collaboration. While none of them immediately calls for a
beyond-SM explanation right now, it is nevertheless interesting to see
whether one can relate these two sets of data following the principle
of Occam's razor, {\em i.e.} by the introduction of a minimal set of
higher-dimensional operators.

We find that this is indeed possible. However, not all operators
invoked in the literature to explain the $R(D^{(*)})$ can do the
job. The situation apparently becomes even more complicated from the
fact that no LFV Higgs coupling can survive if the scalar sector is
SM-like. However, this can be circumvented by postulating the
existence of new degrees of freedom at a higher scale while the
low-energy scalar sector remains completely SM-like. This also leads
to new four-fermion operators which can possibly contribute to $b\to
c\tau\nu$ decays. Arguing that the undetermined nature of the neutrino
flavor allows for the anomaly to be explained in terms of the
muon-neutrino, we relate it, through the $SU(2)_L$ symmetry to the
$\tau\mu$ final state.  While many Lorentz structures, {\em per se.,}
could explain the anomaly(ies), only some survive the stringent limits
imposed by the $Z$ and $B_s$ decays.

We find that it is indeed possible to find a parameter space where
both the anomalies can be successfully explained, with the fit showing
a very marked improvement over the SM. This region is also physically
meaningful in the sense that all the Wilson coefficients for the new
operators are of the order of unity.

This scenario can be tested in a number of ways. First, the $\tau$
polarization, $P_\tau$, can be measured with much improved precision
in future $B$ factories. The SM $\tau$s are all left-chiral, while our
model predicts a large number of right-chiral $\tau$s as well. The
second way is to investigate the LFV couplings of the Higgs boson in
future electron-positron colliders. As has been shown in
Ref.\ \cite{ILC-LFV}, the International Linear Collider can have a
reach one order of magnitude better than the LHC. As for which models
can produce such effective operators, we leave that to the model
builders.

{\em Acknowledgements} -- 
D.C.\ acknowledges partial support from the European Union's Horizon 2020
research and innovation program under Marie Sk{\l}odowska-Curie grant
No 674896. A.K.\ acknowledges the Council for
Scientific and Industrial Research, Government of India, for a
research grant. He also thanks the Physics Department of IIT,
Guwahati, for hospitality where a part of this work was completed.


\begin{thebibliography}{99}

\bibitem{Heeck} 
A comprehensive list of all LFV processes and their limits in present and future 
experiments can be found in, {\em e.g.}, \\
M.~Lindner, M.~Platscher and F.~S.~Queiroz,
  ``A Call for New Physics : The Muon Anomalous Magnetic Moment and Lepton Flavor Violation,''
  arXiv:1610.06587 [hep-ph];\\
  J.~Heeck,
  ``Interpretation of Lepton Flavor Violation,'' Phys.\ Rev.\ D {\bf 95}, 015022 (2017)
  [arXiv:1610.07623 [hep-ph]].
  
\bibitem{1502.07400} 
V.~Khachatryan {\it et al.} [CMS Collaboration],
  ``Search for Lepton-Flavour-Violating Decays of the Higgs Boson,''
  Phys.\ Lett.\ B {\bf 749}, 337 (2015)
  [arXiv:1502.07400 [hep-ex]].
  
  \bibitem{1607.03561}
  V.~Khachatryan {\it et al.} [CMS Collaboration],
  ``Search for lepton flavour violating decays of the Higgs boson to 
  $e \tau$ and $e \mu$ in proton-proton collisions at $\sqrt{s}=8$ TeV,''
  Phys.\ Lett.\ B {\bf 763}, 472 (2016) 
  [arXiv:1607.03561 [hep-ex]];\\
  G.~Aad {\it et al.} [ATLAS Collaboration],
  ``Search for a Heavy Neutral Particle Decaying to $e\mu$, $e\tau$, or $\mu\tau$ in $pp$ Collisions at $\sqrt{s}=8$ TeV with the ATLAS Detector,''
ÊÊPhys.\ Rev.\ Lett.\  {\bf 115}, 031801 (2015)
ÊÊ
ÊÊ[arXiv:1503.04430 [hep-ex]].
ÊÊ
  
\bibitem{1604.07730} 
G.~Aad {\it et al.} [ATLAS Collaboration],
  ``Search for lepton-flavour-violating decays of the Higgs and $Z$ bosons with the ATLAS detector,''
  arXiv:1604.07730 [hep-ex];\\
  G.~Aad {\it et al.} [ATLAS Collaboration],
  ``Search for lepton-flavour-violating $H\to\mu\tau$ decays of the Higgs boson with the ATLAS detector,''
ÊÊJHEP {\bf 1511}, 211 (2015)
ÊÊ
ÊÊ[arXiv:1508.03372 [hep-ex]].
ÊÊ
  
\bibitem{Lees:2013uzd}
  J.~P.~Lees {\it et al.} [BaBar Collaboration],
  ``Measurement of an Excess of $\bar{B} \to D^{(*)}\tau^- \bar{\nu}_\tau$ Decays and Implications for Charged Higgs Bosons,''
  Phys.\ Rev.\ D {\bf 88}, 072012 (2013)
  [arXiv:1303.0571 [hep-ex]].

\bibitem{Huschle:2015rga}
  M.~Huschle {\it et al.} [Belle Collaboration],
  ``Measurement of the branching ratio of $\bar{B} \to D^{(\ast)} \tau^- \bar{\nu}_\tau$ relative to $\bar{B} \to D^{(\ast)} \ell^- \bar{\nu}_\ell$ decays with hadronic tagging at Belle,''
  Phys.\ Rev.\ D {\bf 92}, 072014 (2015)
  [arXiv:1507.03233 [hep-ex]].

\bibitem{Abdesselam:2016cgx}
  A.~Abdesselam {\it et al.} [Belle Collaboration],
  ``Measurement of the branching ratio of $\bar{B}^0 \rightarrow D^{*+} \tau^- \bar{\nu}_{\tau}$ relative to $\bar{B}^0 \rightarrow D^{*+} \ell^- \bar{\nu}_{\ell}$ decays with a semileptonic tagging method,''
  arXiv:1603.06711 [hep-ex].

\bibitem{Aaij:2015yra}
  R.~Aaij {\it et al.} [LHCb Collaboration],
  ``Measurement of the ratio of branching fractions $\mathcal{B}(\bar{B}^0 \to D^{*+}\tau^{-}\bar{\nu}_{\tau})/\mathcal{B}(\bar{B}^0 \to D^{*+}\mu^{-}\bar{\nu}_{\mu})$,''
  Phys.\ Rev.\ Lett.\  {\bf 115}, no. 11, 111803 (2015)
  [Phys.\ Rev.\ Lett.\  {\bf 115}, no. 15, 159901 (2015)]
  [arXiv:1506.08614 [hep-ex]].



\bibitem{Hirose:2016wfn} 
  S.~Hirose {\it et al.} [Belle Collaboration],
  ``Measurement of the $\tau$ lepton polarization and $R(D^*)$ in the decay $\bar{B} \to D^* \tau^- \bar{\nu}_\tau$,''
  arXiv:1612.00529 [hep-ex].
  
 \bibitem{crivellin-heeck}
 A.~Crivellin, J.~Heeck and P.~Stoffer,
  ``A perturbed lepton-specific two-Higgs-doublet model facing experimental hints for physics beyond the Standard Model,''
  Phys.\ Rev.\ Lett.\  {\bf 116}, no. 8, 081801 (2016)
  [arXiv:1507.07567 [hep-ph]].
  
   \bibitem{girish}
 D.~Das, C.~Hati, G.~Kumar and N.~Mahajan,
  ``Towards a unified explanation of $R_{D^{(\ast)}}$, $R_{K}$ and $(g-2)_{\mu}$ anomalies in a left-right model with leptoquarks,''
  Phys.\ Rev.\ D {\bf 94}, 055034 (2016)
  [arXiv:1605.06313 [hep-ph]];\\
  D.~Becirevic, S.~Fajfer, N.~Kosnik and O.~Sumensari,
  ``Leptoquark model to explain the $B$-physics anomalies, $R_K$ and $R_D$,''
  Phys.\ Rev.\ D {\bf 94}, 115021 (2016) [arXiv:1608.08501 [hep-ph]].
    
\bibitem{Bhattacharya:2016} 
  S.~Bhattacharya, S.~Nandi and S.~K.~Patra,
  ``Looking for possible new physics in $B\to D^{(\ast)}\tau\nu_{\tau}$ in light of recent data,''
  arXiv:1611.04605 [hep-ph].

\bibitem{byakti}
D.~Bardhan, P.~Byakti and D.~Ghosh,
  ``A closer look at the $R_D$ and $R_{D^*}$ anomalies,''
  arXiv:1610.03038 [hep-ph].
  
  \bibitem{crivellin}
  J.~Heeck, M.~Holthausen, W.~Rodejohann and Y.~Shimizu,
  ``Higgs  $\to\mu\tau$ in Abelian and non-Abelian flavor symmetry models,''
  Nucl.\ Phys.\ B {\bf 896}, 281 (2015)
  [arXiv:1412.3671 [hep-ph]].\\
    A.~Crivellin, G.~D'Ambrosio and J.~Heeck,
  ``Explaining $h\to\mu^\pm\tau^\mp$, $B\to K^* \mu^+\mu^-$ and $B\to K \mu^+\mu^-/B\to K e^+e^-$ in a two-Higgs-doublet model with gauged $L_\mu-L_\tau$,''
  Phys.\ Rev.\ Lett.\  {\bf 114}, 151801 (2015)
  [arXiv:1501.00993 [hep-ph]];\\
  D.~Aristizabal Sierra and A.~Vicente,
  ``Explaining the CMS Higgs flavor violating decay excess,''
  Phys.\ Rev.\ D {\bf 90}, no. 11, 115004 (2014)
  [arXiv:1409.7690 [hep-ph]];\\
I.~Dorsner, S.~Fajfer, A.~Greljo, J.~F.~Kamenik, N.~Kosnik and I.~Nisandzic,
  ``New Physics Models Facing Lepton Flavor Violating Higgs Decays at the Percent Level,''
  JHEP {\bf 1506}, 108 (2015)
  [arXiv:1502.07784 [hep-ph]];\\
 I.~de Medeiros Varzielas, O.~Fischer and V.~Maurer,
  ``$ {\mathbb{A}}_4 $ symmetry at colliders and in the universe,''
  JHEP {\bf 1508}, 080 (2015)
  [arXiv:1504.03955 [hep-ph]];\\
  W.~Altmannshofer, S.~Gori, A.~L.~Kagan, L.~Silvestrini and J.~Zupan,
  ``Uncovering Mass Generation Through Higgs Flavor Violation,''
  Phys.\ Rev.\ D {\bf 93}, no. 3, 031301 (2016)
  [arXiv:1507.07927 [hep-ph]].
  
  
  
  
  \bibitem{dipankar}
  D.~Das and A.~Kundu,
  ``Two hidden scalars around 125 GeV and h→μτ,''
  Phys.\ Rev.\ D {\bf 92}, no. 1, 015009 (2015)
  [arXiv:1504.01125 [hep-ph]];\\
  M.~Sher and K.~Thrasher,
  ``Flavor Changing Leptonic Decays of Heavy Higgs Bosons,''
  Phys.\ Rev.\ D {\bf 93}, no. 5, 055021 (2016)
  [arXiv:1601.03973 [hep-ph]].
  
  \bibitem{willey}
  R.~S.~Willey and H.~L.~Yu,
  ``Neutral Higgs Boson From Decays of Heavy Flavored Mesons,''
  Phys.\ Rev.\ D {\bf 26}, 3086 (1982).
  
  
  \bibitem{bird}
  C.~Bird, P.~Jackson, R.~V.~Kowalewski and M.~Pospelov,
  ``Dark matter particle productions in $b\to s$ transitions with missing energy,"
  Phys.\ Rev.\ Lett.\  {\bf 93}, 201803 (2004)
  [hep-ph/0401195].
  
  
  \bibitem{ellis}
  G.~Blankenburg, J.~Ellis and G.~Isidori,
  ``Flavour-Changing Decays of a 125 GeV Higgs-like Particle,''
  Phys.\ Lett.\ B {\bf 712}, 386 (2012)
  [arXiv:1202.5704 [hep-ph]].
  
  
  \bibitem{harnik}
  R.~Harnik, J.~Kopp and J.~Zupan,
  ``Flavor Violating Higgs Decays,''
  JHEP {\bf 1303}, 026 (2013)
  [arXiv:1209.1397 [hep-ph]].
  
  
  

\bibitem{cgk}
D.~Choudhury, D.~K.~Ghosh and A.~Kundu,
  ``B decay anomalies in an effective theory,''
  Phys.\ Rev.\ D {\bf 86}, 114037 (2012)
  [arXiv:1210.5076 [hep-ph]].
  
  \bibitem{pilaftsis}
  A.~Pilaftsis,
  ``Lepton flavor nonconservation in H0 decays,''
  Phys.\ Lett.\ B {\bf 285}, 68 (1992);\\
  C.~Alvarado, R.~M.~Capdevilla, A.~Delgado and A.~Martin,
  ``Minimal Models of Loop-Induced Higgs Lepton Flavor Violation,''
  Phys.\ Rev.\ D {\bf 94}, 075010 (2016)
  [arXiv:1602.08506 [hep-ph]].
  
  \bibitem{sakaki}
  Y.~Sakaki, M.~Tanaka, A.~Tayduganov and R.~Watanabe,
  ``Probing New Physics with $q^2$ distributions in $\bar{B} \to D^{(*)} \tau \bar\nu$,''
  Phys.\ Rev.\ D {\bf 91}, no. 11, 114028 (2015)
  [arXiv:1412.3761 [hep-ph]].
  
  \bibitem{srimoy}
  S.~Bhattacharya, S.~Nandi and S.~K.~Patra,
  ``Optimal-observable analysis of possible new physics in $B\to D^{(\ast)}\tau\nu_{\tau}$,''
  Phys.\ Rev.\ D {\bf 93}, no. 3, 034011 (2016)
  [arXiv:1509.07259 [hep-ph]].
  
\bibitem{Choudhury:2013jta} 
  D.~Choudhury, A.~Kundu and P.~Saha,
  ``Z-pole observables in an effective theory,''
  Phys.\ Rev.\ D {\bf 89}, 013002 (2014)
  [arXiv:1305.7199 [hep-ph]].
  
\bibitem{Caprini:1997mu} 
  I.~Caprini, L.~Lellouch and M.~Neubert,
  ``Dispersive bounds on the shape of $\bar{B}\to D^{(*)} \ell\bar\nu$ form-factors,''
  Nucl.\ Phys.\ B {\bf 530}, 153 (1998)
  doi:10.1016/S0550-3213(98)00350-2
  [hep-ph/9712417].
  
  
\bibitem{Amhis:2014hma} 
  Y.~Amhis {\it et al.} [Heavy Flavor Averaging Group (HFAG) Collaboration],
  ``Averages of $b$-hadron, $c$-hadron, and $\tau$-lepton properties as of summer 2014,''
  arXiv:1412.7515 [hep-ex].
  
  
\bibitem{hfag}
The URL is \url{https://www.slac.stanford.edu/xorg/hfag/semi}.

\bibitem{Na:2015kha}
  H.~Na {\it et al.} [HPQCD Collaboration],
  ``$B\to D\ell\nu$ form factors at nonzero recoil and extraction of $|V_{cb}|$,''
  Phys.\ Rev.\ D {\bf 92}, 054510 (2015)
  [arXiv:1505.03925 [hep-lat]].


\bibitem{Kamenik:2008tj}
  J.~F.~Kamenik and F.~Mescia,
  ``$B\to D\tau\nu$ Branching Ratios: Opportunity for Lattice QCD and Hadron Colliders,''
  Phys.\ Rev.\ D {\bf 78}, 014003 (2008)
  [arXiv:0802.3790 [hep-ph]].

 \bibitem{Bozek:2010xy} 
  A.~Bozek {\it et al.} [Belle Collaboration],
  ``Observation of $B^+\to {\bar{D}}^{*0} \tau^+ \nu_\tau$ and Evidence for 
  $B^+ \to {\bar{D}}^0 \tau^+\nu_\tau$ at Belle,''
  Phys.\ Rev.\ D {\bf 82}, 072005 (2010)
  [arXiv:1005.2302 [hep-ex]].
  
\bibitem{b2dlnu}
 C.~Patrignani {\it et al.} [Particle Data Group],
 Chin.\ Phys.\ C {\bf 40}, 100001 (2016).
 
 \bibitem{grinstein}
  R.~Alonso, B.~Grinstein and J.~Martin Camalich,
  ``The lifetime of the $B_c^-$ meson and the anomalies in $B\to D^{(*)}\tau\nu$,''
  arXiv:1611.06676 [hep-ph].


 \bibitem{ILC-LFV} 
 S.~Banerjee, B.~Bhattacherjee, M.~Mitra and M.~Spannowsky,
  ``The Lepton Flavour Violating Higgs Decays at the HL-LHC and the ILC,''
  JHEP {\bf 1607}, 059 (2016)
  [arXiv:1603.05952 [hep-ph]];\\
I.~Chakraborty, A.~Datta and A.~Kundu,
  ``Lepton flavor violating Higgs boson decay ${\boldsymbol{h}} \rightarrow \mu \tau $ at the ILC,''
  J.\ Phys.\ G {\bf 43}, no. 12, 125001 (2016)
  [arXiv:1603.06681 [hep-ph]].
  \end{thebibliography}
\end{document}